\documentclass[prd,aps,preprint,nofootinbib,preprintnumbers]{revtex4}
\usepackage{dcolumn}
\usepackage{bm}
\usepackage{indentfirst}
\usepackage{graphicx}

\newcommand{\bi}{\begin{itemize}}
\newcommand{\ei}{\end{itemize}}
\newcommand{\be}{\begin{equation}}
\newcommand{\ee}{\end{equation}}
\newcommand{\ba}{\begin{eqnarray}}
\newcommand{\ea}{\end{eqnarray}}

\begin{document}


\title{Lepton Number Violating Processes Mediated by Majorana
Neutrinos at Hadron Colliders}

\author{Sergey Kovalenko}
\author{Zhun Lu}
\author{Ivan Schmidt}\affiliation{Departamento de F\'\i sica,\\
Universidad T\'ecnica Federico Santa Mar\'\i a,\\ Casilla 110-V,
Valpara\'\i so, Chile \\and Center of Subatomic Physics, Valpara\'\i
so, Chile}

\begin{abstract}
We study the Lepton number violating like-sign dilepton processes
$h_1 h_2 \to l^\pm l^{\prime \pm} jj  X$ and $h_1 h_2 \to l^\pm
l^{\prime \pm} W^\pm  X$, mediated by heavy GeV scale Majorana
neutrinos. We focus on the resonantly enhanced contributions with a
nearly on-mass-shell  Majorana neutrino in the $s$-channel.  We
study the constraints on like-sign dilepton production at the
Tevatron and the LHC on the basis of the existing experimental
limits on the masses of heavy neutrinos and their mixings $U_{\alpha
N}$ with $\alpha = \nu_{e}, \nu_{\mu}, \nu_{\tau}$. Special
attention is paid to the constraints from neutrinoless double beta
decay. We note that searches for like-sign $e^{\pm} e^{\pm}$ events
at Tevatron and LHC may shed light on CP-violation in neutrino
sector. We also discuss the conditions under which it is possible to
extract individual constraints on the mixing matrix elements in a
model independent way.
\end{abstract}

 \maketitle

\section{Introduction}

At present there are no doubts that
neutrinos are massive particles mixing with each other. Moreover,
according to the neutrino oscillation experiments,  their masses are
extremely small while mixing is nearly maximal.  In this respect
neutrinos drastically defer from all other known particles, and this
difference represents one of the pressing problems for theory.
The famous see-saw mechanism showed the way to possible solutions of
this problem via introduction of very heavy Majorana particles mixed
with ordinary neutrinos, bringing in the observable mass
hierarchy  between the light neutrinos and other fermions. These
heavy particles could be heavy Majorana neutrinos, as in the
original formulation of see-saw, or other new particles such as
neutralinos, heavy Majorana particles mixing with neutrinos in the
framework of SUSY models with R-parity violation.  The new heavy
particles may lead to observable effects beyond the light neutrino
sector, since they could manifest themselves indirectly via their
virtual contribution to processes involving ordinary particles.
If the masses of some of these new particles are not extremely
large, lying within the reach of the running of forthcoming
experiments, they can be searched for directly among the products of
colliding particles.

Here we consider an extended see-saw scenario, including $n$ species
of SM singlet right-handed neutrinos
$\nu^{\,\prime}_{Rj}=(\nu^{\,\prime}_{R1},...\nu^{\,\prime}_{Rn})$,
besides the three left-handed weak doublet neutrinos
$\nu^{\,\prime}_{Li} =
(\nu^{\,\prime}_{Le},\nu^{\,\prime}_{L\mu},\nu^{\,\prime}_{L\tau})$.
The general mass term for this set of fields can be written as \ba -
\frac{1}{2} \overline{\nu^{\,\prime}} {\cal M}^{(\nu)} \nu^{\,\prime
c} + \mbox{h.c.} & = & - \frac{1}{2} (\bar\nu^{\,\prime}_{_L},
\overline{\nu_{_R}^{\,\prime c}}) \left(\begin{array}{cc}
{\cal M}_L & {\cal M}_D \\
{\cal M}^T_D  & {\cal M}_R \end{array}\right) \left(\begin{array}{c}
\nu_{_L}^{\,\prime c} \\
\nu^{\,\prime}_{_R}\end{array}\right) + \mbox{h.c.} \\
&=&-\frac{1}{2} (\sum_{i=1}^{3} m_{\nu_i} \overline{\nu^c_i}\nu_i
+\sum_{j=1}^{n} m_{\nu_j} \overline{\nu^c_j}\nu_j  )+ \mbox{h.c.}
\ea Here ${\cal M}_L, {\cal M}_R$ are $3\times 3$ and $n\times n$
symmetric Majorana mass matrices, and ${M}_D$ is a $3\times n$ Dirac
type matrix. Rotating the neutrino mass matrix to the diagonal form by a unitary
transformation
 \ba U^T {\cal M}^{(\nu)}U =
\textrm{Diag}\{m_{\nu 1}, \cdots ,m_{\nu_{3+n}}\}
\label{mass-eigen}
\ea
 we end up with $3+n$
Majorana neutrinos with masses $m_{v_1}, \cdots, m_{v_{3+n}}$. In this scenario the light
neutrinos have the mass scale $\mathcal{M}_D^2/\mathcal{M}_R$, and
as a result there should be also heavy Majorana neutrinos ($N$)
with mass scale $\mathcal{M}_R$. Due to the fact that Majorana
neutrinos are also their anti-neutrinos, they can produce lepton
number ($L$) violation by two units. One of the decisive processes
for probing the Majorana nature of neutrinos is neutrinoless double
beta ($0\nu\beta\beta$) decay
which has been extensively
studied in the literature~\cite{Elliott:2004hr,Benes:2005hn}.
Majorana neutrinos may also resonantly contribute to meson \cite{Dib:2000wm}  and $\tau$ decays \cite{Gribanov:2001vv}.

Another potential
process to look for Majorana neutrinos is like-sign dilepton
production $h_1 h_2 \to l^\pm l^{\prime \pm} W^\mp X$ at hadron
colliders~\cite{Almeida:2000pz,Panella:2001wq,Han:2006ip,Han:2009ip}. In this
paper we give a detailed analysis of like-sign dilepton
production processes, evaluating the cross section of
$h_1 h_2 \to l^\pm l^{\prime \pm} jj X$ and $h_1 h_2 \to l^\pm l^{\prime \pm}W^\mp X$.

The existing very stringent constraint from $0\nu\beta\beta$
experiments on the inverse effective mass $U^{2}_{eN}/M_{N}$ of a
heavy Majorana neutrino N is frequently treated as the limit, which
makes unrealistic the experimental observation of the above
processes with $e^{\pm}e^{\pm}$ pair in the final state. However, we
note that CP-violating Majorana phases, if present in the neutrino
mixing matrix elements $U_{e\alpha}$,  are able to significantly
soften or even completely evade the $0\nu\beta\beta$-constraints on
the $e^{\pm}e^{\pm}$-process production.  We point out that this can
be true even in the presence of only one heavy Majorana neutrino,
due to the interference of heavy-light neutrino contribution to
$0\nu\beta\beta$-decay.

The paper is organized as follows. In the next section we present our approach to calculation of the cross sections of
the  $h_1 h_2 \to l^\pm l^{\prime \pm} jj X$ and $h_1 h_2 \to l^\pm l^{\prime \pm}W^\mp X$ processes.
In Sec. III we calculate the decay width of the heavy Majorana neutrino entering in the above cross sections.
Sec. IV is devoted to a discussion of the existing limits on the parameters of heavy Majorana neutrino and their impact
on the prospects for searches for like-sign dileptons at Tevatron and LHC. Here we comment on implications of the
$0\nu\beta\beta$-decay constraints and CP-violation in the neutrino sector and discuss the possibility of extracting
the heavy-light neutrino mixing matrix elements $U_{eN}, U_{\mu N}, U_{\tau N}$ in a model independent way.

\section{theoretical framework}

\begin{figure}[t]
\begin{center}
%
\scalebox{0.75}{\includegraphics*[112,546][292,693]{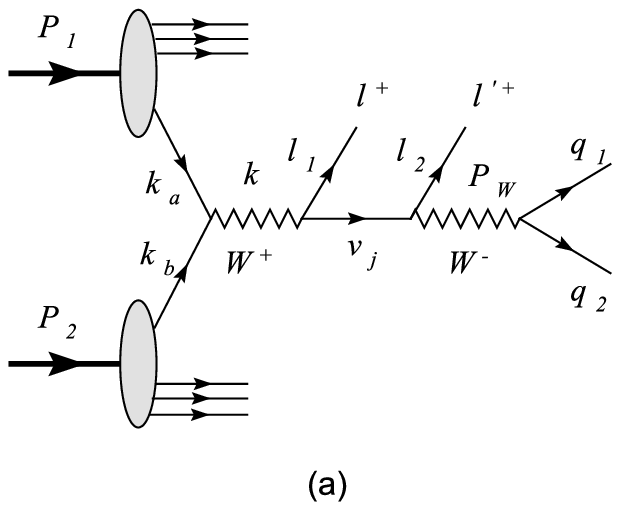}
\includegraphics*[101,546][238,698]{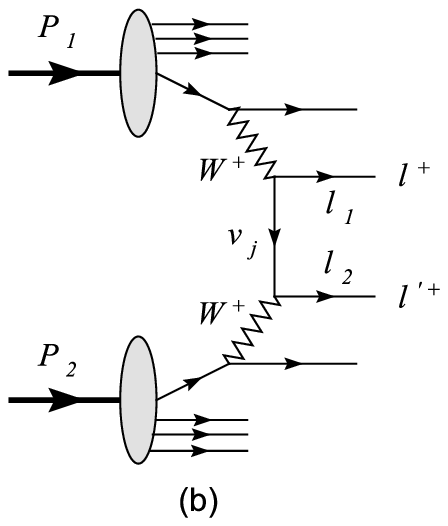}}
%
%
\caption{\small
(a) resonant  and (b) non-resonsnt Majorana neutrino contributions to \mbox{$h_1 (P_1) + h_2 (P_2) \to l^+
(l_1) + l^{\prime +} (l_2) + X $} process. In the diagrams we indicate the momenta of the corresponding particles.
}
 \label{diag}
\end{center}
\end{figure}

The processes we study are like-sign dilepton inclusive production in
high energy $pp$ or $p\bar{p}$ collisions
\begin{eqnarray} \label{Proc-def}
h_1  \, (P_1) + h_2 \, (P_2) \to l^{\pm} (l_1) + l^{\prime \pm}(l_2) + jj(W^\mp)
+ X,
\end{eqnarray}
where $jj$ and  $X$ denotes two quark jets and undetected hadronic states. The two leptons
$l$ and $l^\prime$ can have the same or different lepton flavors. In these processes the total lepton number $L$
is violated in two units $\Delta L = 2$. If one assumes the
existence of a Majorana neutrino $N$, the process can be
realized through the diagrams shown in Fig.~\ref{diag}.  We focus on  the resonantly enhanced diagram
Fig.~\ref{diag}(a) with the nearly on mass-shell neutrino in the kinematical region of the studied processes.



 In the diagram Fig.~\ref{diag}(a)  there is a $W$ boson in the final state, which
decays to hadron or lepton states. Here we will consider
the situation in which $W^-$ decays to two quark jets, as shown in
Fig.~\ref{diag}a. According to the factorization theorem, we can
express the cross section of the process as
\begin{eqnarray}
 \sigma^{h_1 + h_2 \to l^++ l^{\prime +}   + X} = \sum_{k,l}\int dx_1 \int
dx_2 \, \{q_{a/h_1}(x_a) \, \bar{q}_{b/h_2}(x_b) \times  \sigma^{q_a
\bar{q}_b \to l^+ l^{\prime +} q_k \bar{q}_l} + (x_a \leftrightarrow
x_b)\},\label{cross1}
\end{eqnarray}
here $q_{a/h_1}(x_a)$ and $\bar{q}_{b/h_2}(x_b)$ denote the
densities of partons inside each hadron, and $d\sigma^{q_a\bar{q}_b
\to l^+ l^{\prime +} q_k \bar{q}_l}$ is the differential
cross section of the partonic subprocess
\begin{eqnarray}\label{subproccess-1}
q_a (k_a) + \bar{q}_b
(k_b) \to  (l_1) + l^\prime + W (P_W) \to l (l_1) + l^\prime + q_k
(q_1)+\bar{q}_l (q_2)
\end{eqnarray}
which has the form
\begin{eqnarray}
d \sigma^{q_a \bar{q}_b \to l^+  l^{\prime +} q_k \bar{q}_l }& =&
\frac{1}{2\hat{s}}\left |\mathcal{M}^{q_a \bar{q}_b \to l^+
l^{\prime +} q_k \bar{q}_l} \right |^2(2\pi)^4
\delta^4(k_a+k_b-l_1-l_2-q_1-q_2) \nonumber \\
&\times&
 \frac{d^3l_1}{(2\pi)^3
2E_1} \frac{d^3l_2}{(2\pi)^3 2E_2} \frac{d^3q_1}{(2\pi)^3
2E_{q_1}}\frac{d^3q_2}{(2\pi)^3 2E_{q_2}}.\label{subpro4}
\end{eqnarray}
The next step is to calculate the amplitude squared
$|\mathcal{M}|^2$, which can be written as
\begin{eqnarray}
|\mathcal{M}|^2=16G_F^4M_W^8\times
\frac{(|V_{ud}|^2u(x_a)\bar{d}(x_b)+|V_{us}|^2u(x_a)\bar{s}(x_b))
H^{\mu\nu, \alpha\beta} \, L_{\mu\nu,
\alpha\beta}}{((\hat{s}-M_W^2)^2+\Gamma_W^2
M_W^2)((P_W^2-M_W^2)^2+\Gamma_W^2 M_W^2)} +(x_a \leftrightarrow
x_b),
\end{eqnarray}
with the tensors having the following forms:
\begin{eqnarray}
H^{\mu\nu, \alpha\beta} & = & 16(k_a^{\mu} k_b^{\nu}+k_a^{\nu}
k_b^{\mu} -g^{\mu \nu} k_a \cdot k_b) (q_1^{\alpha}
q_2^{\beta}+q_1^{\beta} q_2^{\alpha} -g^{\mu \nu} q_1
\cdot q_2) \\
L^{\mu\nu, \alpha\beta} &=&  \left [ \sum_N \frac{ m_N
U_{lN}U_{l'N}}{[(k-l_1)^2-m_N^2]} \bar{v}(l_1)\gamma^\mu \gamma^\alpha P_R
v^c(l_2)-(l_1 \leftrightarrow l_2)\right ]\nonumber
\\
& \times & \left [ \sum_N \frac{ m_N U_{lN}U_{l'N}}{[(k-l_1)^2-m_N^2]}
\bar{v}(l_1)\gamma^\nu \gamma^\beta P_R v^c(l_2)-(l_1
\leftrightarrow l_2) \right ]^\dag, \label{lmunu4}
\end{eqnarray}
The contraction of the above two tensors yields
\begin{eqnarray}
H^{\mu\nu, \alpha\beta} \, L_{\mu\nu, \alpha\beta} &=& 128\sum_N
\frac{ m_N^2 |U_{l N}U_{l'N}|^2}{[(k-l_1)^2-m_N^2]^2} \left ((l_1\cdot
k_b)(l_2\cdot q_2)(k_a\cdot q_1)+(l_1\cdot k_b)(l_2\cdot q_1)(k_a
\cdot q_2) \right .\nonumber \\&+& \left .(l_1\cdot k_a)(l_2\cdot
q_2)(k_b\cdot q_1)+ (l_1\cdot k_b)(l_2\cdot q_2)(k_b\cdot q_2)\right
)+~\textrm{ interference term}. \label{contraction1}
\end{eqnarray}
We have checked that the interference term gives a very small
contribution compared to the first term in (\ref{contraction1}), and
therefore we ignore it in the following calculations. The 4-body
phase space in Eq.~(\ref{subpro4}) can be rewritten as
\begin{eqnarray}
\Phi_4 & = &\delta^4(k_a+k_b-l_1-l_2-P_W)
 \frac{d^3l_1}{(2\pi)^3
2E_{l_1}} \frac{d^3l_2}{(2\pi)^3 2E_{l_2}} \frac{d^3P_W}{(2\pi)^3
2E_{W}}\nonumber\\
& \times &\delta^4(P_W-q_1-q_2) \frac{d^3q_1}{(2\pi)^3
2E_{q_1}}\frac{d^3q_2}{(2\pi)^3 2E_{q_2}}(2\pi)^3
dP_W^2,\label{4body}
\end{eqnarray}
where $P_W=q_1 +q_2$ denotes the momentum of the final virtual $W$
boson. In the c.m. frame of $q_1$ and $q_2$, the two body
phase space can be expressed as
\begin{eqnarray}
\delta^4(P_W-q_1-q_2) \frac{d^3q_1}{(2\pi)^3
2E_{q_1}}\frac{d^3q_2}{(2\pi)^3 2E_{q_2}} \rightarrow
\frac{\lambda^{1/2}(P_W^2,m_{q_1}^2,m_{q_2}^2)}{8(2\pi)^6  P_W^2}\,
d\Omega,
\end{eqnarray}
where $d\Omega=d\cos\theta_1 d\phi_1$ is the solid angle of
$\boldsymbol{q}_1$, and the form of
$\lambda(P_W^2,m_{q_1}^2,m_{q_2}^2)$ is given in Eq.~(\ref{gram3}).
In the case $P_W^2 \gg m_{q_1}^2$ and $m_{q_2}^2$,
$\lambda^{1/2}(P_W^2,m_{q_1}^2,m_{q_2}^2) \sim P_W^2$. After
integrating over $d\Omega$, Eq.~(\ref{contraction1}) turns to
\begin{eqnarray}
\int d\Omega \, H^{\mu\nu, \alpha\beta} \, L_{\mu\nu, \alpha\beta}
&=& \frac{128 \pi}{3}\sum_N
\frac{ m_N^2 |U_{lN}U_{l'N}|^2}{[(k-l_1)^2-m_N^2]^2} P_W^2((l_1 \cdot k_b)
(2E_{l_2}E_{k_a}+l_2 \cdot k_a )
\nonumber \\
 &+&  (l_1 \cdot k_a)  (2E_{l_2}E_{k_b}+l_2 \cdot k_b )),
\label{contraction2}
\end{eqnarray}
and the four-body phase space in Eq.~(\ref{4body}) is reduced to a
three-body phase space.

\begin{figure}[b]
\begin{center}
%
\scalebox{1.0}{\includegraphics*{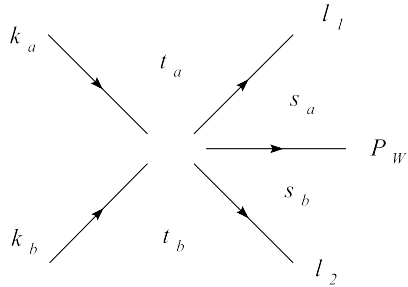}}
%
%
\caption{Definition of the invariant variables.}
 \label{invariant}
\end{center}
\end{figure}

As shown in Fig.~\ref{invariant}, it is convenient to use the
following Lorentz invariants in order to express the kinematical
variables:
\begin{eqnarray}
\hat{s} &=& (k_a+k_b)^2 = x_a x_b s, ~~~s_a = (l_1+P_W)^2,\nonumber\\
s_b &=& (l_2+P_W)^2,~~~t_a = (k_a-l_2)^2, ~~~ t_b = (k_b-l_1)^2.
\label{t_b-dif}
\end{eqnarray}
With the above invariant variables, we can apply the phase space
transformation given in Ref.~\cite{Kajantie:1969hd} to rewrite the
cross section given in Eq.~(\ref{cross1}) as
\begin{eqnarray}
\sigma &=& a\, \int d x_a \int d x_b \int_{s_a^-}^{s_a^+}  d s_a
\int_{t_a^{-}}^{t_a^{+}} d t_a \int_{s_b^-}^{s_b^+} d s_b
\int_{t_b^-}^{t_b^+} d t_b \int d P_W^2 \nonumber \\
& \times & \frac{\Theta (-\Delta_4)}{\hat{s}^2\sqrt{-\Delta_4}}\,
\mathcal{M}(x_a,x_b,s_a,s_b,t_a,t_b,P_W^2), \label{cross3}
\end{eqnarray}
whith $a=2G_F^4M_W^8 /(3(2\pi)^6)$, and
\begin{eqnarray}
\mathcal{M}(x_a,x_b,s_a,s_b,t_a,t_b) &=&
\frac{|V_{ud}|^2u(x_a)\bar{d}(x_b)+|V_{us}|^2u(x_a)\bar{s}(x_b) +
(x_a \leftrightarrow x_b)}{((\hat{s}-M_W^2)^2+\Gamma_W^2
M_W^2)((P_W^2-M_W^2)^2+\Gamma_W^2 M_W^2)}
\label{amplitude1} \\
& \times & \left |\sum_N \frac{ m_N U_{lN}U_{l'N}}{s_b-m_N^2} \right |^2
\mathcal{M}_1(\hat{s},s_a,s_b,t_a,t_b,P_W^2) + \textrm{interference
term}, \nonumber
\end{eqnarray}
where
\begin{eqnarray}
 \mathcal{M}_1(\hat{s},s_a,s_b,t_a,t_b,P_W^2)&=&(\hat{s}-s_b+t_a-m_a^2)((s_b-P_W^2-m_{l_2}^2)(s_a+t_a-t_b-m_{l_1}^2)
           \nonumber\\&+&P_W^2(\hat{s}-s_a+t_b-m_b^2))
          -(t_a-m_a^2-m_{l_1}^2)((s_b-P_W^2-m_{l_2}^2)\nonumber\\
 &\times&     (s_b-t_a+t_b-m_{l_2}^2)-P_W^2(t_b-m_b^2-m_{l_2}^2)),
\end{eqnarray}
The phase space can be constrained to the physical region by the
following inequalities for the Gram determinants
\begin{equation}
\Delta_3(p_1,p_2,p_3) \geq 0, ~~~~~~~~~\Delta_4(p_1,p_2,p_3,p_4)
\leq 0, \label{gramdet}
\end{equation}
where $\{p_i\}$ denotes any subset of  momenta $\{k_a, k_b, l_1,
l_2, P_W\}$. The resulting integration limits $s_{a/b}^-, s_{a/b}^+,
t_{a/b}^-$ and $t_{a/b}^+$ are given in Appendix.

%
%
%

If there exists one Majorana neutrino with the mass $m_N$
is in the region $s_b^-<m_N^2<s_b^+$, the integrand shown in
Eq.~(\ref{cross3}) has a pole at $s_{b}= m_N^2$.
Introducing a decay width $\Gamma_{N}$ for $N$ through the
substitution $m_N \to m_N-(i/2)\Gamma_{N}$, the Majorana neutrino
propagator can be written as $1/(s-m_N^2+i m_N\Gamma_{N})$. As we
will show in the next section, $\Gamma_{N}\ll m_{N}$ for $m_{N}$ in  range we are studying in the present paper.
Therefore, the cross section can be written as
\begin{eqnarray}\label{ll+jj}
\sigma^{h_1h_2\to l^+  l^{\prime +}  X}(m_N) & \approx & a\,\pi \int
d x_a \int d x_b \int_{s_a^-}^{s_a^+}  d s_a \int_{t_a^-}^{t_a^+} d
t_a  \int_{t_b^-}^{t_b^+} d t_b\int P_W^2 \frac{\Theta
(-\Delta_4)}{\hat{s}^2 \sqrt{-\Delta_4}}\nonumber \\
 &\times & \frac{|V_{ud}|^2u(x_a)\bar{d}(x_b)+|V_{us}|^2u(x_a)\bar{s}(x_b) +
(x_a \leftrightarrow x_b)}{((\hat{s}-M_W^2)^2+\Gamma_W^2
M_W^2)((P_W^2-M_W^2)^2+\Gamma_W^2 M_W^2)} \nonumber \\
 &\times &  \frac{ m_N
\left |U_{ lN }U_{ l'N }\right |^2}{\Gamma_{N}}
\mathcal{M}_1(\hat{s},s_a,m_N^2,t_a,t_b,P_W^2),\label{csfinal}
\end{eqnarray}
 To arrive
at the above equation we have used the identity:
\begin{eqnarray}
\int_{s_b^-}^{s_b^+} ds_b
\frac{f(s_b)}{(s_b-m_N^2)^2+m_N^2\Gamma_{N}^2} = \pi
\frac{f(m_N)}{m_N\Gamma_{N}},~~ ~~ s_b^-<m_N^2<s_b^+  ~~
\textrm{and} ~~ \Gamma_{N}<< m_N. \label{resonance}
\end{eqnarray}
Therefore, with $\Gamma_{N}$ in the denominator, the cross section
can be strongly enhanced through the resonant production of a
Majorana neutrino. If there are several Majorana neutrinos in the
range $s_b^-<m_N^2<s_b^+$, there should be a sum over $N_{i}$ in
Eq.~(\ref{csfinal}).

%
%
%
%

In the case $m_N> M_W$, one can detect the $W$ boson together with
the lepton pair $l^+ l^{\prime +}$ in the final state. Therefore we
also give the formula to calculate the cross section of the process
$h_1 + h_2 \to l^+ + l^{\prime +} + W^- + X $,
\begin{eqnarray}\label{ll+W}
\sigma^{h_1h_2\to l^+  l^{\prime +}  W^-  X}(m_N) & \approx
&a^{\prime} \, \pi \int d x_a \int d x_b \int_{s_a^-}^{s_a^+}  d s_a
\int_{t_a^-}^{t_a^+} d t_a \int_{t_b^-}^{t_b^+}d t_b \frac{\Theta
(-\Delta_4)}{\hat{s}^2 \sqrt{-\Delta_4}}\nonumber \\
&\times &
\frac{|V_{ud}|^2u(x_1)\bar{d}(x_2)+|V_{us}|^2u(x_1)\bar{s}(x_2) +
(x_1 \leftrightarrow x_2)}{(\hat{s}-M_W^2)^2+\Gamma_W^2 M_W^2}
\nonumber \\
 &\times &
\frac{ m_N  \left |U_{lN}U_{l'N}\right |^2}{\Gamma_{N}}
\mathcal{M}_1^{\prime}(\hat{s},s_a,m_N^2,t_a,t_b),\label{csfinal2}
\end{eqnarray}
where
\begin{eqnarray}
\mathcal{M}_1^{\prime} (\hat{s},s_a,s_b,t_a,t_b) & = &
 (s_a+t_a-t_b-m_{l_1}^2)(\hat{s}-s_b+t_a-m_a^2)(s_b-M_W^2-m_{l_2}^2)/M_W^2\nonumber \\
    & - &   (t_a-m_a^2-m_{l_1}^2)(s_b-t_a+t_b-m_3^2)(s_b-M_W^2-m_{l_2}^2)/M_W^2 \nonumber \\
     & + &
     (m_b^2+m_{l_2}^2-t_b)(m_a^2+m_{l_1}^2-t_a) \nonumber\\
    &+&(\hat{s}-s_a+t_b-m_b^2)(\hat{s}-s_b+t_a-m_a^2),\label{amplitude2}
\end{eqnarray}
and $a^\prime = 2\sqrt{2}G_F^3M_W^6/(6(2\pi)^4)$.

\section{Neutrino decay width}

As shown in Eq.~(\ref{csfinal}) and (\ref{csfinal2}), the
cross sections of the process $h_1 h_2 \to l^+ l^{\prime +} X$,
%
and $h_1 h_2 \to l^\pm l^{\prime \pm}W^\mp X$
mediated by the resonant production of Majorana neutrino depend on the neutrino
decay width $\Gamma_N$.
%
For $m_N \ll M_W$ the decay width $\Gamma_{N}$
receives contributions from
leptonic and semi-leptonic channels
 \ba \label{channels-1}
 ~~~~N\rightarrow l_1\bar{l}_2\nu,~\nu_i(\nu_i^c)l\bar l,~ l^\pm q_1\bar{q}_2,~ \nu_i (\nu_i^c)\, q \bar{q}.
 \ea
  We approximate the semi-leptonic decays by inclusive quark-antiquark pair production.  In such an approach \cite{Gribanov:2001vv}
  based on the  Bloom-Gilman duality \cite{BloomGilman}
%
the total decay width of the heavy Neutrino $N$ to hadrons is reproduced
in average with an accuracy
sufficient for our estimates.  One of the advantages of this approach is that it does not  require knowledge of meson
masses $M_{\cal{M}}$ and decay constants $F_{\cal{M}}$ necessary for the calculation of $N\rightarrow l \cal{M}$ partial
widths which then are summed up in order to derive the total decay width of heavy neutrino $N$ in the channel-by-channel
approach \cite{Han:2009ip}.  For heavy mesons these parameters are purely known and may introduce a significant uncertainty.

The leading order decay rates for the channels listed in (\ref{channels-1}) can be found for the leptonic and semileptonic decays
in Refs.~\cite{Gribanov:2001vv, Han:2009ip,Panella:2001wq} and \cite{Gribanov:2001vv} respectively.
%
We summarize the corresponding formulas neglecting lepton and quark masses.  In this approximation
the partial decay widths of Majorana neutrino in the region $m_{N}\leq M_{W}$ are
\ba \label{decayrates}
%
\Gamma(N\rightarrow l_1\bar{l}_2\nu_{l_{2}} )&=& |U_{l_1 N}|^2
\frac{G_F^2}{192\pi^3} m_N^5 F_{W}(m_{N})
 \equiv |U_{l_1 N}|^2 \Gamma^{l\nu}_{1}, \\
 \label{N-llnu}
%
\Gamma(N\rightarrow \nu_{l_{1}} l_{2}\bar l_{2})&=& |U_{l_{1}N}|^2 \frac{G_F^2}{96 \pi^3}m_N^5F_{Z}(m_{N})\times \\
\nonumber
&\times&[(g_{L}^{l})^{2}+(g_{R}^{l})^{2} +
g_{L}^{l}g_{R}^{l}+ \delta_{l_{1}l_{2}}(g_{R}^{l} + 2 g_{L}^{l})] \equiv |U_{l_{1}N}|^2\Gamma^{l\nu}_{2},\\
 \label{N-3nu}
%
\sum_{l_{2}=e,\mu,\tau} \Gamma(N\rightarrow \nu_{l_{1}}\bar \nu_{l_{2}}\nu_{l_{2}})&=& |U_{l_{1}N}|^2 \frac{G_F^2}{96 \pi^3}m_N^5
F_{Z}(m_{N}) \equiv  |U_{l_{1}N}|^2 \Gamma^{3\nu},\\
 \label{N-lud}
%
\Gamma(N\to l_{1}^{-} u\bar{d})&=&|U_{l_{1}N}|^2\ (|V_{u d}|^{2}+|V_{u s}|^{2}+ |V_{cs}|^{2} )\frac{G_F^2}{64\pi^3}m_N^5
F_{W}(m_{N})\equiv \\
\nonumber
&\equiv& |U_{l_{1}N}|^2\Gamma^{lud},\\
 \label{N-nuqq}
\Gamma(N\to \nu_i\, q\bar{q})\ \ &=&|U_{l_{1}N}|^2\frac{G_F^2}{32\pi^3}m_N^5F_{Z}(m_{N})\times \\
\nonumber
&\times&[(g_{L}^{q})^{2}+(g_{R}^{q})^{2} +
g_{L}^{q}g_{R}^{q}]
=|U_{l_{1}N}|^2\Gamma^{\nu q},
 \ea
In Eq. (\ref{N-lud}) we neglected the small charged current $cb$ contribution. Here  $g^{l}_{L}=-1/2 + \sin^2\theta_W,  g^{l}_{R} = \sin^2\theta_W, g^{u}_{L}= 1/2 - (2/3) \sin^2\theta_W,
g^{u}_{R}= -(2/3)\sin^2\theta_W,  g^{d}_{L}= -1/2 + (1/3) \sin^2\theta_W, g^{d}_{R}= (1/3)\sin^2\theta_W$
are the SM neutral current lepton and quark couplings where $l=e,\mu,\tau$ and $u=u,c,t$;   $d=d,s,b$.

In Eqs. (\ref{decayrates})-(\ref{N-nuqq}) the function
 \begin{equation}
F_{B}(m_N) = 6M_B^4\int_0^{m_N^2} d s_1 \int_0^{m_N^2-s_1} d s_2
\frac{2s_1 m_N^2-2s_1 s_2-s_1^2+s_2 m_N^2-s_2^2}{m_{N}^8((s_2 -
M_B^2)^2 + \Gamma_B^2M_B^2)}.
\end{equation}
takes into account a propagator effect of W and Z-bozon exchange and introduce a significant correction of a factor $\sim 3$ for
$m_{N}\sim M_{W}, M_{Z}$. We denote $B=W,Z$. Here $M_{W} = 80.4$GeV,
$\Gamma_{W} = 2.14$GeV and $M_{Z}=91.2$GeV, $\Gamma_{Z} = 2.5$GeV are the mass and full decay width of W and Z-bosones
respectively.

Therefore, in the case $m_N\leq M_W$, the total decay rate is
%
%
\ba
\Gamma_{N}(m_N\leq m_W) &\approx& \left( \sum_{l_1= e, \mu,\tau}|U_{lN}|^2\right) \left [2 \ \Gamma^{l}_{1} +2\  \Gamma^{lud}
+\Gamma^{l \nu}_{2} + \Gamma^{3\nu}+ \sum_{q=u,d,s,c,b}\Gamma^{\nu q}\right ].
 \ea
 %
%
%
%
%
The factor 2 in the first two terms is due to Majorana nature of heavy neutrino N which can decay into two non-equivalent charge conjugate final states.
%
%
For $m_N>M_W$ the total decay of heavy neutrino is determined by the decay channels:
\begin{eqnarray}\label{DR-2}
N \rightarrow l^{\mp}W^{\mp}, \ \ \  N \rightarrow \nu Z, \ \ \ N \rightarrow \nu H^{0}.
\end{eqnarray}
The corresponding partial
widths of Majorana neutrino are given by~\cite{Han:2009ip,Pilaftsis:1991ug}:
\begin{eqnarray}
\Gamma (N \to l^{\pm}W^{\mp}) & = &|U_{lN}|^2  \frac{G_F}{8\sqrt{2}\pi}
m^3_{N} (1+\frac{2M^2_W}{m^2_N})(1-\frac{M^2_W}{m^2_N})^2
\theta (m_N-M_W)=|U_{lN}|^2 \Gamma^{(lW)},\nonumber\\
\\
\Gamma (N \to \nu_l (\nu_l^c) Z^0) & = &
|U_{lN}|^{2}\frac{G_F}{8\sqrt{2}\pi} m^3_j
(1+\frac{2M^2_Z}{m^2_N})(1-\frac{M^2_Z}{m^2_N})^2 \theta
(m_N-M_Z)=|U_{lN}|^2\Gamma^{(\nu Z)},\\
\Gamma (N  \to \nu_i (\nu_i^c) H^0) &=&|U_{lN}|^2 \frac{G_F}{8\sqrt{2}\pi}
m^3_j (1-\frac{M^2_H}{m^2_N})^2 \theta (m_N-M_H) =|U_{lN}|^2\Gamma^{(\nu H)} .
\end{eqnarray} 
In the same way we can write the total decay width of  a Majorana
neutrino at $m_N>M_W$ as
%
%
%
\ba \label{DR-GG}
\Gamma_{N}(m_N>M_W) \approx \left(\sum_{l=e, \mu, \tau}|U_{lN}|^2\right)(2 \Gamma^{W}+\Gamma^{Z}+
\Gamma^{H})
\ea

The above formulas we use in our analysis of the like-sign dilepton production. Numerically
$\Gamma_{N}\ll m_N$  holds for $1 \mbox{GeV} \leq m_{N} \leq 1\mbox{TeV}$ and, therefore,
Eqs.~(\ref{csfinal}) and (\ref{csfinal2}) are good
approximations for the cross sections of the process $h_1 h_2 \to l^+ l^{\prime +} jj(W) X$.


\section{Limits on heavy Majorana neutrino sector}
\label{Limits}
The remaining ingredient, which determines in the most crucial way the event rate of the like-sign dilepton production,
is the heavy Majorana neutrino masses and their mixing with $\nu_{e}, \nu_{\mu}, \nu_{\tau}$.
In the literature there are various limits on $M_{N}$ and $U_{\alpha N}$ for
$\alpha=e, \mu, \tau$, extracted from direct experimental searches \cite{PDG-N}
for these particles in a wide region of masses, from the non-observation of lepton number violating decays,
as well as from precision measurements of certain observable quantities \cite{Almeida:2000}.
In the present context we are interested in these limits for the mass range $M_{N}\geq 1 \textrm{GeV}$.
From the global fit of the electroweak precision measurements, including LEP data, it was found \cite{Almeida:2000}:
\begin{eqnarray}\label{LEP-lim}
|U_{eN}|^{2}\leq 0.0052, \ \ \  |U_{\mu N}|^{2}\leq 0.0001, \ \ \  |U_{\tau N}|^{2}\leq 0.01
\end{eqnarray}
The LEP data on direct searches of heavy leptons ~\cite{LEP-searches}
shows that
\begin{eqnarray}\label{LEP-search}
|U_{lN}|^2 \leq 10^{-4}-10^{-5}\ \ \ \mbox{for}\ \ \  3 \mbox{GeV}\leq M_N < 80  \mbox{GeV}
\end{eqnarray}
where $l=e, \mu, \tau$.
%

Neutrinoless double beta decay ($0\nu\beta\beta$) is known to be a sensitive probe of
Majorana neutrino masses and mixing.
Presently the best the experimental lower
bound~\cite{KlapdorKleingrothaus:2000sn} on the
$0\nu\beta\beta$-decay half life was obtained for ${}^{76}$Ge:
\begin{equation}\label{H-M}
T^{0\nu}_{1/2}(^{76}Ge) \geq 1.9\times 10^{25}\mbox{yrs}.
\end{equation}
In Ref. ~\cite{Benes:2005hn} this bound was used to constrain
the contribution of Majorana neutrinos of arbitrary mass. In a good approximation this constrain reads:
\begin{eqnarray}\label{dbd-1}
\sum_{k}
\frac{\left |U_{e k}\right|^{2} e^{i\alpha_k} m_{\nu k}}{m_{\nu k}^{2} + q_0^{2}}\leq 5\times 10^{-8} \,\textrm{GeV}^{-1}.
\end{eqnarray}
with $q_0 = 105$ MeV. Here $\alpha_k$ denotes the Majorana CP-phase of the Majorana neutrino state $\nu_k$ of mass $m_{\nu k}$. For the light-heavy neutrino scenario one has:
\begin{eqnarray}\label{dbd-2}
\sum_{N=\textrm{heavy}}
\frac{\left |U_{e N}\right|^{2} }{M_N} e^{i\alpha_N}
+ q_0^{-2}\sum_{i=\textrm{light}}
\left |U_{e i}\right|^{2} e^{i\alpha_i} m_{\nu i}
\leq 5\times 10^{-8} \,\textrm{GeV}^{-1}.
\end{eqnarray}
where $M_N\gg q_0$ and $m_{\nu i}\ll q_0$.
%
%
This stringent constraint, applied to each term separately, seems to
make unrealistic observation of like-sign dielectrons
(positrons). However, the possibility of CP violation in the
neutrino sector should be taken into account.
If there are more than one heavy neutrino state $N_k$,  with
different CP phases $\alpha_N$, then different terms in the first sum of (\ref{dbd-2}) may
compensate each other, reducing the individual
$0\nu\beta\beta$-constrain on each of them.
%
Note that even with the presence of only one heavy neutrino there
may happen the above mentioned compensation between the heavy $N$ and
the three light neutrinos $\nu_i$, since they all coherently contribute to
$0\nu\beta\beta$-decay. Thus searching for like-sign dielectron
(positron) events may provide evidence for CP violation in the
neutrino sector. In case of observation of these events at a rate
larger than that stemming from the $0\nu\beta\beta$-limit applied to
each term in (\ref{dbd-2}), this would point to  CP-violation in the
neutrino sector.

\begin{figure}[t]
\begin{center}
\scalebox{0.38}{\includegraphics{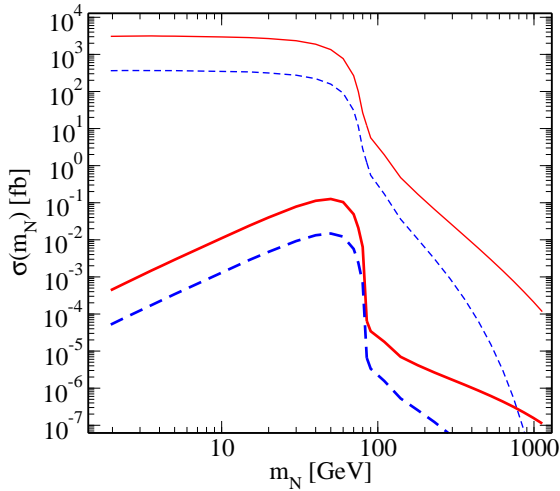}}
%
\caption{\small
The cross section of $p + p(\bar{p})\to e^\pm e^\pm jj + X$ mediated
by a heavy Majorana neutrino, at the LHC (solid line) and Tevatron
(dashed line), as a function of $m_N$. The thick lines correspond to
the constraint from the $0\nu\beta\beta$-decay experimental data and
the thin lines are calculated from the fixed mixing elements
$|U_{\tau N}|^2 \sim |U_{\mu N}|^2 \sim |U_{e N}|^2=10^{-3}$.}
 \label{lhc1}
\end{center}
\end{figure}
\begin{figure}[t]
\begin{center}
\scalebox{0.38}{\includegraphics[9pt,32pt][552pt,508pt]{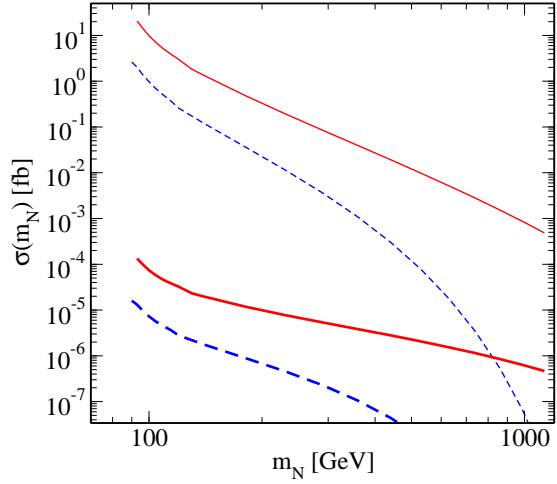}}
%
%
\caption{\small Same as in Fig.~\ref{lhc1}, but for $p + p(\bar{p})\to e^\pm e^\pm W^\mp  X$.}
 \label{real-w}
\end{center}
\end{figure}

With this in mind we have calculated the maximum of the
$p +p(\bar{p}) \to e^\pm e^\pm jjX$ and $p +p(\bar{p}) \to e^\pm e^\pm W^{\mp}X$ cross sections at the Tevatron and
LHC, consistent with the $0\nu\beta\beta$-limit (\ref{dbd-2}), assuming
no CP violation in the neutrino sector. The results are shown by the
thick lines in Figs~\ref{lhc1} and \ref{real-w}.
In the calculation we used the limits
\begin{eqnarray}\nonumber
&&\frac{|U_{e N}|^2}{M_N}\leq 5 \times 10^{-8}\mbox{GeV}^{-1},\ \ \
|U_{\mu N}|^2 \leq 10^{-4},\\
\label{lim-1}
&&|U_{\tau N}|^2 \leq 10^{-4}\ \ \mbox{for} \ \
3 \mbox{GeV} \leq m_N \leq 80 \mbox{GeV},\\
\nonumber
&&|U_{\tau N}|^2 \leq 10^{-2}\ \ \mbox{for} \ \
80 \mbox{GeV} > m_N.
\end{eqnarray}
For comparison, we also give the cross section calculated with the
assumption  $|U_{\tau N}|^2 \sim |U_{\mu N}|^2 \sim
|U_{e N}|^2=10^{-3}$, shown in Figs.~\ref{lhc1} and \ref{real-w} by thin lines.

\begin{figure}[t]
\begin{center}
\scalebox{0.4}{\includegraphics[4pt,73pt][592pt,510pt]{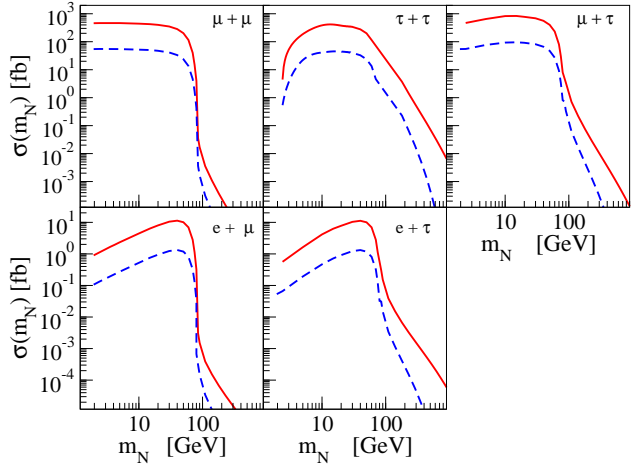}}
%
%
\caption{\small The bounds on the cross section of $p +p(\bar{p})
\to l^\pm l^{\prime \pm} jjX$ at the LHC (solid line) and Tevatron
(dashed line) as a function of $m_N$ for $l l^\prime =
 \mu\mu,\tau\tau,\mu\tau,e\mu$ and $e\tau$.}
 \label{other-flavors}
\end{center}
\end{figure}

\begin{figure}[t]
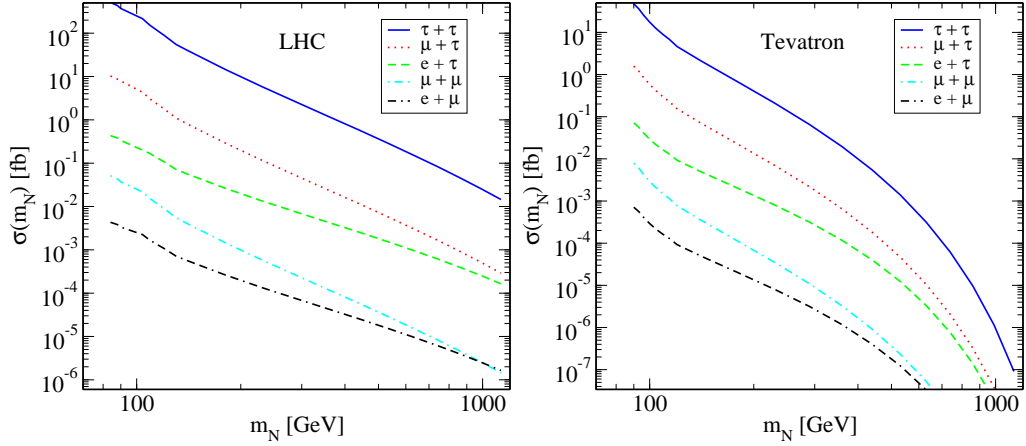

\begin{center}
\scalebox{0.35}{\includegraphics{LHC-real-W.eps}}
\scalebox{0.35}{\includegraphics{Tevatron-real-W.eps}}
%
%
\caption{\small The bounds on the
cross section of $p +p(\bar{p}) \to l^\pm l^{\prime \pm}  W^\mp X$ at the LHC
(left figure) and Tevatron (right figure) as a function of $m_N$ for $l l^\prime =
 \mu\mu,\tau\tau,\mu\tau,e\mu$ and $e\tau$.}
 \label{W-other-flavor}
\end{center}
\end{figure}

With the same constraints (\ref{lim-1})
we calculated bounds on
$p +p(\bar{p}) \to l^\pm l^{\prime \pm} jjX$ and $p+ p(\bar{p}) \to l^\pm l^{\prime \pm}  W^{\mp}X$
cross sections for other lepton flavors, at the LHC and Tevatron, and the results are plotted in Figs.~\ref{other-flavors}
and \ref{W-other-flavor}.

In Tables.~\ref{events-LHC} and ~\ref{events-Tevatron} we show the upper limits for $p + p(\bar{p})
\to l^\pm l^{\prime \pm}  jjX$ event number for all possible lepton flavors, calculated with the constraints in Eq.
(\ref{lim-1}), at the LHC with integrated luminosity 10 fb$^{-1}$, and at the Tevatron with integrated luminosity
2fb$^{-1}$.

The numbers given in these tables, except for the like-sign $\mu \mu$-production, do not take into account event selection cuts
necessary to suppress the background and which may also dramatically affect theoretical predictions for the signal.
In the present paper we do not address this issue. In order to evaluate an impact of the experimental cuts on our results
we use the results of the recent analysis \cite{Han:2006ip,Han:2009ip} of the like-sign $\mu \mu$ selection criteria for
the case of the LHC and TEVATRON experiments.
In Ref. \cite{Han:2006ip} there were found sets of the cuts which permit efficient suppression of the background and confident
selection of $\mu^{\pm} \mu^{\pm}$-signal.  In Tables I and II we indicate in brackets the number of $\mu^{\pm} \mu^{\pm}$-events
after applying these cuts for the LHC and Tevatron experiments. As seen from these tables, the regions
%
where there is still a chance to observe $\mu^{\pm} \mu^{\pm}$-production with statistical significance are
$20\mbox{GeV}\leq m_{N} \leq 70\mbox{GeV}$ for LHC and $20\mbox{GeV}\leq m_{N} \leq 50\mbox{GeV}$ for Tevatron.
As to the other lepton flavors there exist in the literature studies
(see, {\it e.g.} Refs.  \cite{Almeida:2000pz, Panella:2001wq} ) of the corresponding selection cuts, although
not so detailed as in Ref. \cite{Han:2009ip}. Unfortunately the effect of these cuts on our results cannot be
directly recognized and requires a special analysis to be done elsewhere. Roughly, according to
Refs.  \cite{Almeida:2000pz, Panella:2001wq}, the cuts reduce the like-sign dilepton event by
an order of magnitude for $m_{N}\leq 100$GeV,  leaving practically unchanged the region of higher $m_{N}$ values.

The following comment is in order here. Notice that because of the
involvement of all the mixing matrix elements $U_{e N}, U_{\mu N},
U_{\tau N}$ in the like-sign dilepton production cross section
(\ref{ll+jj})-(\ref{ll+W}), one needs to rely on various scenarios
relating these matrix elements, in order extract limits on them from
experimental data. These additional not well motivated assumptions
decrease reliability of such results. Naturally it would be
desirable to get the possibility of extracting individual limits on
each of the three mixing matrix  elements $U_{e N}, U_{\mu N},
U_{\tau N}$ without any ad hoc assumptions on their relative sizes.
Such an extraction is possible in the studied case of production of
like-sign dileptons by simultaneous search for events with different
lepton flavors. As follows from (\ref{ll+jj})-(\ref{ll+W})
%
\begin{eqnarray}\label{indep}
\sigma^{h_{1}h_{2}\rightarrow l e}+\sigma^{h_{1}h_{2}\rightarrow l \mu} + \sigma^{h_{1}h_{2}\rightarrow l \tau} =
|U_{l N}|^{2}
\end{eqnarray}
Thus, searching for one flavor diagonal and two flavor non-diagonal
processes and establishing upper bounds on each cross section within
the same kinematical domain, one can set an upper limit on the
corresponding mixing matrix element $|U_{l N}|$. Therefore searches for like-sign dileptons with all the lepton flavors are required for the model independent extraction of each mixing matrix element $|U_{l N}|$.  An experimentally challenging point is searching for the events with $\tau$'s since they involve missing momentum.
In this respect it is worth mentioning a method for reconstructing $\tau$-momenta on the basis of the analysis of the isolated charged tracks from $\tau$-decays proposed in Ref. \cite{Han:2009ip}. The authors have argued their method to be a promising basis for future searches for like-sign dileptons with one or two $\tau$'s.

\begin{table}
\begin{tabular}{c|cccccc}
\hline  $m_N$ (GeV) &$ee$&$e\mu$ & $e\tau$ &$\mu\mu$&$\mu\tau$&$\tau\tau$\\
\hline
\hline $ ~10~$ &$ ~0~$& $ ~44~$ &$ ~ 41~$& $ ~4.4\cdot10^3~$ ($0$)& $ ~8.3\cdot10^3~$&$ ~3.8\cdot10^3~$\\
\hline  20 &0& 80 &78&$ 4.0\cdot10^3$  ($4$)& $ 7.8 \cdot 10^3$ & $ 3.8 \cdot 10^3$\\
\hline  30 &1& 105& 104& $3.5\cdot10^3$ ($10$)& $6.9 \cdot 10^3$& $3.4\cdot10^3$ \\
\hline  50 &1&101 & 101& $2.0 \cdot 10^3$ ($10^{2}$)& $4.0 \cdot 10^3$&$2.0 \cdot 10^3$ \\
 \hline 70 &0& 28 & 29 & 401(20)&801 & $1.0 \cdot 10^3$\\
 \hline 90 &0& 0 & 2 & 0& 34&877\\
  \hline 110&0 & 0 & 0 &0 & 7&690\\
  \hline
\end{tabular}\caption{ Maximum event number of $p p \to l^\pm l^\pm jjX$
at the LHC (10 fb$^{-1}$ data)   consistent with the limits in
Eq.~(\ref{lim-1}).
Indicated in  brackets are typical number of events after the event selection cuts of Refs. \cite{Han:2006ip,Han:2009ip}.
}\label{events-LHC}
\end{table}

\begin{table}
\begin{tabular}{c|cccccc}
\hline  $m_N$ (GeV) &$ee$&$e\mu$ & $e\tau$ &$\mu\mu$&$\mu\tau$&$\tau\tau$\\
\hline
\hline$ 10 $&$ ~~0~~ $&$ ~~1~~$ &$ ~~1~~$&$ ~~104$ (2)&$ ~~ 187~~$&$ ~~82~~$\\
\hline  20 &0&2 &2&95 (4)& 183&89\\
\hline  30 &0& 2& 2& 82 (5)& 163& 80 \\
\hline  50 &0&2 & 2& 47 (6)& 95&47 \\
 \hline 70 &0& 1 & 1 & 9 (0)&18 & 9\\
 \hline 90 &0& 0 & 2 & 0& 1&8\\
  \hline 110 &0& 0 & 0 &0 &0&6\\
  \hline
\end{tabular}\caption{ Same as Table.~\ref{events-LHC}, but for $p \bar{p} \to l^\pm l^\pm jjX$ at
the Tevatron (2 fb$^{-1}$ data).}\label{events-Tevatron}
\end{table}


\section{Summary and Conclusions}

We have studied like-sign $\Delta L=2$  dilepton
inclusive production mediated by heavy Majorana
neutrinos. We focussed on the dominant mechanism via resonant
Majorana neutrino production.
We applied the existing limits on heavy Majorana neutrino mass $m_{N}$ and its mixing with active flavors
$\nu_e, \nu_{\mu}, \nu_{\tau}$,
in order to predict maximal like-sign dilepton production rates for $p +p(\bar{p}) \to l^\pm l^{\prime \pm} jjX$ and
$p +p(\bar{p}) \to l^\pm l^{\prime \pm}  W^\mp X$, at the Tevatron
and the LHC.
Special emphasis has been made on the stringent limit on the heavy Majorana neutrino
derived from $0\nu\beta\beta$-decay experiments.
Our results displayed in Figs. \ref{lhc1} - \ref{W-other-flavor} and  Tables I, II  have been obtained assuming
no CP-violation in the neutrino sector.

As seen from Tables I and II, we expect no good prospects for observation of like-sign $ee$ events from
 the studied reactions  neither at the Tevatron, with 2 fb$^{-1}$ integrated luminosity, nor at the LHC, with an
 integrated luminosity of 10\,fb$^{-1}$. This is a direct consequence of very stringent $0\nu\beta\beta$-decay limits.
With an increasing integrated luminosity
100\,fb$^{-1}$ at the LHC, the like-sign $ee$ production
can reach a few events in the region \mbox{$20\ \mbox{GeV} \le m_N \le 60 $ GeV}.
For the
$p + p(\bar{p}) \to e^\pm e^\pm W^\mp X$ process, in which a real $W$
boson is detected 100\,fb$^{-1}$ data at the LHC are
sufficient for discovering Majorana neutrino through like-sign $ee$ production.
One of the messages of the present paper is that despite of these discouraging predictions the like-sign $ee$ events are worth searching for
both at Tevatron and LHC.  The point is that any observation of such events in these experiments would point to the CP-violation in the neutrino sector, as explained in sec. \ref{Limits}.

Like-sign dilepton production for other flavors at the LHC and
Tevatron have better prospects. In the case of $\mu^{\pm}\mu^{\pm}$ production, where we were able to estimate the effect of the event
selection cuts, the possibility of their observation is open in the regions
\mbox{$20\ \mbox{GeV}\leq m_{N} \leq 70\ \mbox{GeV}$} for LHC and \mbox{$20\ \ \mbox{GeV}\leq m_{N} \leq 50\ \mbox{GeV}$} for the Tevatron.

We proposed a method of model independent extraction of the heavy Majorana neutrino mixing $U_{lN}$ with the neutrino active flavors, on the basis of searches of certain sets of like-sign dileptons with different lepton flavors. This method will be helpful in reconstructing the structure of heavy neutrino sector.


{\bf Acknowledgements} This work is supported by the PBCT project
ACT-028 ``Center of Subatomic Physics" and CONICYT No.FB 0821.

\appendix
\section{Integration limits for $s_{a/b}$ and $t_{a/b}$ }

The limits of integration over  $s_{a/b}^-, s_{a/b}^+, t_{a/b}^-$
and $t_{a/b}^+$, given in Eqs.~(\ref{csfinal}) and (\ref{csfinal2}),
can be obtained in the following way \cite{Kajantie:1969hd}. The fourth order Gram
determinant has the form
\begin{eqnarray}
 \Delta_4(P_W,l_2,k_a,k_b)=\left|
  \begin{array}{cccc}
    P_W\cdot P_W & P_W\cdot l_2 & P_W\cdot k_a & P_W\cdot k_b \\
    l_2\cdot P_W & l_2\cdot l_2 & l_2\cdot k_a & l_2\cdot k_b \\
    k_a\cdot P_W & k_a\cdot l_2 & k_a\cdot k_a & k_a\cdot k_b \\
    k_b\cdot P_W & k_b\cdot l_2 & k_b\cdot k_a & k_b\cdot k_b \\
  \end{array}
\right|
\end{eqnarray}
 From the definitions in Eq.~(\ref{t_b-dif}), we see that $\Delta_4$
is a function of $\hat{s}, s_a,s_b,t_a$ and $t_b$. Picking $s_b$ as
the innermost integration variable, explicit evaluation of
$\Delta_4$ yields
\begin{eqnarray}
16\Delta_4=as_b^2+bs_b+c = a(s_b-s_b^+)(s_b-s_b^-).
\end{eqnarray}
Therefore the $s_b$ limits are
\begin{eqnarray}
s_b^+ & = & \frac{-b+\sqrt{\Delta}}{2a}, \nonumber\\
s_b^- & = & \frac{c}{ a\, s_b^-}
\end{eqnarray}
The $s_a$-integration limits can be obtained from the requirement
$\Delta >0$. Solving this equation yields
\begin{eqnarray}
s_a^+ & = &
(t_b(\hat{s}+m_b^2-m_a^2)+\hat{s}(m_b^2-m_{l_2}^2)+m_{l_2}^2(m_a^2+m_b^2)
-m_b^2(m_b^2-m_a^2)
\nonumber\\&+&\lambda^{1/2}(\hat{s},m_a^2,m_b^2)\lambda^{1/2}(t_b,m_b^2,m_{l_2}^2))
/(2m_b^2),\nonumber\\
s_a^- & = &
t_b+m_a^2+((t1+m_a^2-m_{l_1}^2)(m_{l_2}^2-t_a-t_b)\nonumber \\
& - &
\lambda^{1/2}(t_a,t_b,P_W^2)\lambda^{1/2}(t_a,m_a^2,m_{l_1}^2))/(2t_a),
\end{eqnarray}
where the third order Gram determinant has the form
\begin{eqnarray}
\lambda(x,y,z)= x^2+y^2+z^2-2xy-2yz-2xz. \label{gram3}
\end{eqnarray}
Defining
\begin{eqnarray}
Q&=&\hat{s}(t_b+m_b^2-m_{l_2}^2)-m_a^2t_b+m_{l_2}^2(m_a^2+m_b^2)-m_b^2(m_b^2+m_{l_1}^2+P_W^2)
\nonumber \\
& + & \lambda^{1/2}(\hat{s},m_a^2,m_b^2)\lambda^{1/2}(t_b,m_b^2,m_{l_2}^2),\\
a^\prime&=&2m_b^2(Q+m_b^2(t_b+m_a^2+m_{l_1}^2+P_W^2))\\
b^\prime&=&Q^2-m_b^4(t_b-m_a^2-P_W^2+m_{l_1}^2)^2
-4m_b^4(m_a^2+m_{l_1}^2)(t_b+P_W^2)\\
c^\prime&=&2m_b^2(Q(m_a^2-m_{l_1}^2)(t_b-P_W^2)\nonumber\\
& + & m_b^2(m_a^2-m_{l_1}^2)^2
(t_b+P_W^2)+m_b^2(t_b-P_W^2)^2(m_a^2+m_{l_1}^2)),
\end{eqnarray}
we can arrive at the integration limit for $t_a$:
\begin{eqnarray}
  t_a^-&=&\frac{-b^\prime-\sqrt{{b^\prime}^2-4a^\prime c^\prime}}{2a^\prime},\\
   t_a^+&=&\frac{c^\prime}{a^\prime t_a^-}.
   \end{eqnarray}
Finally we give the $t_b$-integration limits as
\begin{eqnarray}
t_b^{\pm} & = &
(m_{l_1}+\sqrt{P_W^2})^2+m_a^2-((\hat{s}+m_a^2-m_b^2)(\hat{s}+(m_{l_1}+\sqrt{P_W^2})^2
-m_{l_2}^2) \nonumber
\\
& \pm &(\lambda^{1/2}(\hat{s},m_a^2,m_b^2)\lambda^{1/2}
(\hat{s},(m_{l_1}+\sqrt{P_W^2})^2,m_{l_2}^2)))/(2\hat{s}).
\end{eqnarray}

\end{document}